\documentclass[runningheads]{llncs}
\usepackage{graphicx}

\usepackage{dsfont} 
\usepackage{amsmath}
\usepackage{amssymb}
\usepackage{xcolor}

\begin{document}
%
\title{Two-Stage Deep Learning Framework for Quality Assessment of Left Atrial Late Gadolinium Enhanced MRI Images}
\author{K M Arefeen Sultan$^{1,2}$, Benjamin Orkild$^{1,3,4}$, \\
Alan Morris$^{1}$, Eugene Kholmovski$^{5,6}$, \\
Erik Bieging$^{5,7}$, Eugene Kwan$^{3,4}$, \\
Ravi Ranjan$^{3,4,7}$, Ed DiBella$^{3,5}$, \\
Shireen Elhabian$^{1,2}$}
\institute{
$^1$ Scientific Computing and Imaging Institute, University of Utah, SLC, UT \\
$^2$ Kahlert School of Computing, University of Utah, SLC, UT \\
$^3$ Department of Biomedical Engineering, University of Utah, SLC, UT \\
$^4$ Nora Eccles Harrison Cardiovascular Research and Training Institute, University of Utah, SLC, UT \\
$^5$ Department of Radiology and Imaging Sciences, University of Utah, SLC, UT \\
$^6$ Department of Biomedical Engineering, Johns Hopkins, Baltimore, MD \\
$^7$ Division of Cardiology, University of Utah, SLC, UT}
\titlerunning{Left Atrial IQA with Two-Stage DL Model.}
%
\authorrunning{K. Sultan et al.}
%
\maketitle              
\begin{abstract}

Accurate assessment of left atrial fibrosis in patients with atrial fibrillation relies on high-quality 3D late gadolinium enhancement (LGE) MRI images. However, obtaining such images is challenging due to patient motion, changing breathing patterns, or sub-optimal choice of pulse sequence parameters. Automated assessment of LGE-MRI image diagnostic quality is clinically significant as it would enhance diagnostic accuracy, improve efficiency, ensure standardization, and contributes to better patient outcomes by providing reliable and high-quality LGE-MRI scans for fibrosis quantification and treatment planning. To address this, we propose a two-stage deep-learning approach for automated LGE-MRI image diagnostic quality assessment. The method includes a left atrium detector to focus on relevant regions and a deep network to evaluate diagnostic quality. We explore two training strategies, multi-task learning, and pretraining using contrastive learning, to overcome limited annotated data in medical imaging. Contrastive Learning result shows about $4$\%, and $9$\% improvement in F1-Score and Specificity compared to Multi-Task learning when there's limited data.

\end{abstract}
\section{Introduction}

Atrial fibrillation (AF) is currently the most common cardiac arrhythmia in the United States, with 3 to 5 million people affected, and is expected to affect more than 12 million by 2030 \cite{BAO:Col2013}. It has been shown that atrial fibrosis is closely linked to the development and recurrence of AF disease after treatment \cite{BAO:Elm2015,BAO:Mar2014}. Currently, catheter ablation is a popular treatment for AF, targeting and eliminating the areas of the heart (i.e., \textit{fibrotic} tissues) responsible for irregular electrical signals by creating targeted lesions or \textit{scars} in these regions. Hence, fibrosis quantification plays a crucial role in guiding catheter ablation procedures. However, the rate of success of catheter ablation is relatively low, with over 40\% of patients returning to AF within 1.5 years of ablation \cite{BAO:Ver2015}. Therefore, it is imperative to understand and address the shortcomings of the treatment of AF. 

Late Gadolinium Enhancement (LGE) MRI is a widespread technology used to image and quantify myocardial fibrosis and scarring. LGE-MRI can be performed in atrial fibrillation subjects prior to a catheter ablation treatment to provide the patient's atrial geometry and fibrosis pattern \cite{BAO:Oak2009,BAO:Cai2021}. The specific geometry and fibrosis patterns of patients are derived from the LGE-MRI images, which
can be used for pre-ablation planning or for creating patient-specific simulations \cite{BAO:Lan2022,BAO:McD2012}. However, LGE-MRI images exhibit variability in quality, with diagnostic accuracy affected by factors such as noise, resolution, intensity level, and patient-related characteristics \cite{noisylge1,noisylge2,noisylge3}.
The clinical significance of quality assessment in LGE-MRI scans for fibrosis quantification is important as it enhances diagnostic accuracy, ablation planning, and treatment guidance. By discarding poor-quality images, clinicians can base their decisions on more reliable results. However, manual quality assessment is laborious and prone to errors making it non-scalable. Automating this process can optimize workflows and can save resources. This automation necessitates the identification of image features predictive of diagnostic quality, a task where deep learning can be instrumental. However, the effectiveness of deep learning depends on the availability of a substantial amount of annotated images, as it has a significant appetite for labeled data.
Manual annotation of LGE datasets is a laborious and time-consuming expert-driven task, leading to a scarcity of labeled data. 
In this paper, we propose a two-stage deep-learning approach that is inspired by the mental process of a radiologist manually evaluating the diagnostic quality of an LGE MRI image for fibrosis quantification. The proposed method is specially designed to mitigate the limited training data scenario in LGE-MRI quality assessment, leveraging contrastive learning for pretraining and multi-task learning for regularization. The contributions of this paper are summarized as follows.
\begin{enumerate}
    \item[--] Introducing a segmentation network to identify relevant left atrium slices in the LGE-MRI scan instead of relying on manual selection of the slices.
    \item[--] Leveraging a multi-task learning framework to learn quality assessment of atrial fibrosis for the LGE datasets jointly with identification of the atrial blood pool.
    \item[--] Showcasing the impact of label supervision in contrastive learning to promote learning a discriminative representation in the embedding space.
    \item[--] Benchmarking the effectiveness of the two approaches using a limited labeled dataset.
\end{enumerate}

\section{Related Works}
Several automatic methods have been proposed to assess MRI image quality in various anatomical sites. Xu et al.\cite{semibrainiqa} proposed a mean teacher model with ROI consistency to assess the image quality of fetal brain MRI. Liao et al. \cite{brainiqa} used a 2D U-Net to jointly assess image quality and segment fetal brain MRIs. Although they labeled the slices as only good/bad, the assessment of image quality encompasses several interconnected factors. This method also required manual labeling of each training MRI slice, which is tedious and time-consuming. 

To address the limitation of extensive labeling image datasets, various self-supervised learning methods have been introduced. Chen et al. \cite{simclr} proposed using contrastive loss by maximizing the agreement between augmented views of the same image and minimizing the agreement between different images. These learned representations are then applied to downstream tasks, such as image classification and object detection. While this method leverages a vast number of negative samples to learn useful representations, it still relies on a considerable amount of unlabeled data. Khosla et al. \cite{supconloss} proposed label supervision within contrastive loss allowing for more efficient representation learning within limited labeled data.

\section{Methods}

We introduce a two-stage deep-learning approach that emulates the cognitive process of a radiologist manually assessing the diagnostic quality of LGE MRI images for fibrosis quantification. The method consists of two  stages: (1) left atrium detection stage and (2) quality assessment stage.  Fig. \ref{fig:arch} depicts the proposed two-stage approach.

\subsection{Left Atrium Detection Stage}

The primary objective of the LA detection stage is to identify the specific slices within the LGE MRI scan that contain the left atrium. This information is then utilized in the second stage to focus on the relevant region of interest, disregarding any background artifacts that might otherwise interfere with the quality assessment task. Here, we employ a UNet model \cite{unet} to generate segmentation masks for the left atrial blood pool in the MRI scan. The predicted masks undergo a sigmoid function to transform pixel values within the range of 0 to 1. Subsequently, we apply a threshold parameter, \textit{t}, to determine the minimum probability required for a pixel to be considered as part of the left atria. If any pixel value is larger than the threshold, we classify the corresponding slices as containing the left atrium, which is then utilized in the next stage. We exclude the slices where there is no left atrium detected. 

\subsection{Quality Assessment Stage}

This stage incorporates a deep network that effectively maps the image slices to a diagnostic quality score. 
Here, we assess the effectiveness of two training strategies in addressing the challenge of limited annotated data for the quality assessment task: multi-task learning and pretraining using contrastive learning.
We extract the features from a pre-trained network, specifically ResNet$34$ \cite{resnet}. The pre-trained weights are based on the Imagenet dataset. After that, we project the embedding space to a latent space using three attribute classifier modules. 
Each attribute classifier focuses on a fine-grained attribute that is relevant to the quality assessment task.  

\textbf{Image Quality Attributes:} We propose the myocardium nulling, sharpness, and enhancement of aorta and valve attributes that are clinically relevant to the diagnostic quality of fibrosis assessment of LGE-MRIs. 
\textit{Myocardium nulling} compares the intensity of the left ventricular (LV) myocardium to the left ventricular blood pool. A score of $1$ means the intensity of the LV myocardium is higher than that of the blood pool, while a score of $5$ means the intensity of the LV myocardium is well-nulled and similar to that of the signal-free background. 
\textit{Sharpness} reflects the amount of blurring in the borders of the LA and other anatomical structures. A score of $1$ means there is a severe blurring of the cardiac chambers, while a score of $5$ means the edges of the cardiac chambers are well-defined. %
The third attribute is the \textit{enhancement of fibrous tissue - aorta and valve}. When the wall of the aorta and the cardiac valves show enhancement, this implies the scan also has good quality for detection of fibrosis in the left atrium.

For \textit{fibrosis quality assessment}, experts also score quality of fibrous tissues. A score of $5$ defines a high contrast between enhanced fibrous structures and blood pool, whereas $1$ defines the absence of enhanced fibrous structures.


\begin{table}[h]
\centering
\resizebox{\textwidth}{!}{
\begin{tabular}{|c|c|c|c|}
\hline
                                         & \textbf{Myocardium nulling} & \textbf{Sharpness} & \textbf{Enhancement of aorta and valve} \\ \hline
\textbf{Quality of fibrosis assessment} & 0.74                        & 0.79               & 0.76                                    \\ \hline
\end{tabular}
}
\caption{Pearson coefficient.}
\label{table:correlation}
\end{table}
These three attributes, along with fibrosis assessment in the left atrium, were all given scores from 1-5 by trained observers and were transformed to binary scores: non-diagnostic and diagnostic (see Section 4.1). Next, we discuss the training strategies into 3 subsections, Baseline QA, Multi-Task QA, and pretraining using supervised contrastive learning.
\subsubsection{Baseline QA\\\\}

\textbf{Attribute Classifier Module:} The objective of the attribute classifier submodule is to use image features extracted from the image encoder to classify the three attributes, myocardium nulling, sharpness, and enhancement of aorta and valve, into non-diagnostic and diagnostic. Let these attributes be denoted as a vector $\textbf{a}=[a_{mn}, a_{s}, a_{eat}]$ where $a_{mn}$, $a_{s}$, and $a_{eat}$ denote the score of myocardium nulling, sharpness, and enhancement of aorta and valve, and the score $a_{*} \in \{0,1\}$. These 3 attribute classifiers are trained using BCE loss, described in equation \ref{l_qa}. 

\begin{figure}[h]
\centering
\includegraphics[width=1.0\textwidth]{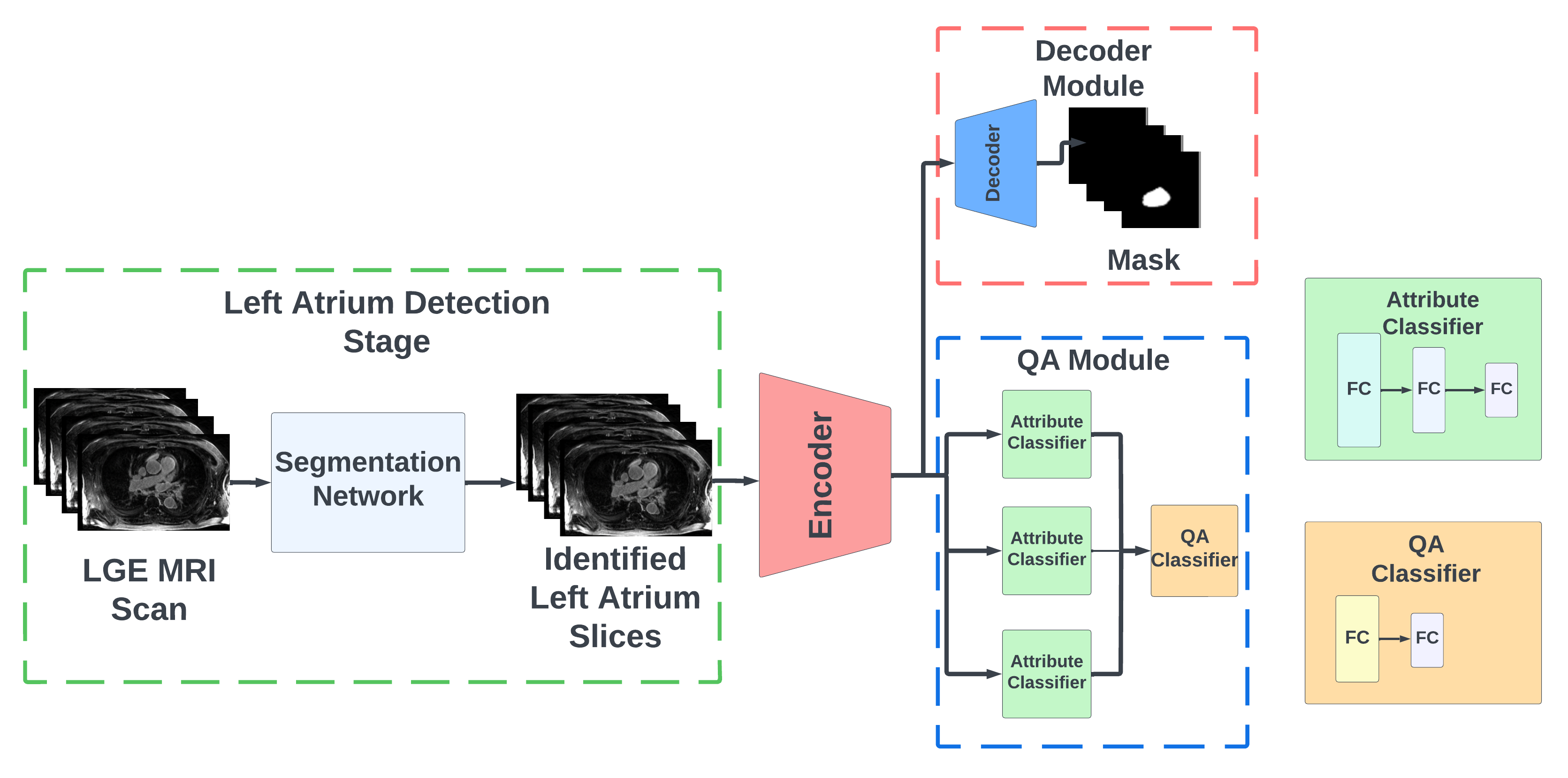}
\caption{Architecture of our model.}
\label{fig:arch}
\end{figure}

\noindent
\textbf{QA module:} The purpose of this module is to predict the quality of fibrosis assessment. Since the $3$ attributes correlate highly with fibrosis assessment as shown in Table \ref{table:correlation}, the module for Quality Assessment (QA), shown in Figure \ref{fig:arch}, concatenates the output of these $3$ attribute classifiers discussed above and then predicts $\hat{y_{qa}}$. 
The network is trained by minimizing a binary cross-entropy (BCE) loss, $\mathcal{L}_{qa}$, which combines the attribute loss and QA loss. \\
Then the supervised loss of Baseline QA is defined by
\begin{equation}
    \mathcal{L}_{qa} = \text{BCE}([a_{mn}, a_{s}, a_{eft}, y_{qa}], [\hat{a_{mn}}, \hat{a_{s}}, \hat{a_{eft}}, \hat{y_{qa}}])
\label{l_qa}
\end{equation}
where $\hat{a_{*}}$, and $\hat{y_{qa}}$ are the prediction of the network.

\subsubsection{Multi-Task QA\\\\}
\textbf{Decoder Module:} The Decoder module is responsible for transforming the embedding space of the encoder into a segmentation mask. The goal of this module is to segment the blood pool, which helps to provide discriminative features for the scoring task. With the segmentation of the blood pool and the QA network, the overall architecture focuses on the area of the left atrium, shown in Figure \ref{fig:hirescam}, where the scoring plays an important role. The segmentation loss is defined by
\begin{equation}
    \mathcal{L}_{seg} = \text{DICE}(\mathbf{M}, \mathbf{\hat{M}})
\end{equation}
where $\mathbf{M}$ is the groundtruth LA segmentation mask, and $\mathbf{\hat{M}}$ is the network-generated mask.\\
Overall, the loss for training the Multi-Task QA network is defined by
\begin{equation}
    \mathcal{L} = \mathcal{L}_{qa} + \mathcal{L}_{seg}
\end{equation}

\subsubsection{Pre-training using supervised contrastive learning\\\\}
Since we have limited labeled data, we run another experiment of pretraining the encoder by utilizing supervised contrastive learning \cite{supconloss}. The motivation behind this approach is to enhance our model's representation learning capabilities so that the same class representation comes closer and pushes the representations of different classes apart. The loss is defined by:
\begin{equation}
    \mathcal{L}_{\text{sup}} = \sum_{i \in I} \frac{-1}{|P(i)|} \sum_{p \in P(i)} \log \frac{\exp(z_i \cdot z_p / \tau)}{\sum_{a=1}^{2N}\mathds{1}_{[a \neq i]} \exp(z_i \cdot z_a / \tau)}
\end{equation}
Here, $P(i)$ is the set of all positives in the augmented view batch corresponding to the anchor $(i)$. For each anchor $i$, there is $1$ positive pair and $2(N-1)$ negative pairs. $z_{*}$ denotes the projected embedding space of the encoder. For our objective, we optimize the loss function below
\begin{equation}
    \mathcal{L}_{s}^{*} = \mathcal{L}_{sup_{mn}} + \mathcal{L}_{sup_s} + \mathcal{L}_{sup_{eat}} + \mathcal{L}_{sup_{qa}} 
\end{equation}
where $\mathcal{L}_{sup_{mn}}$,  $\mathcal{L}_{sup_s}$, $\mathcal{L}_{sup_{eat}}$, and $\mathcal{L}_{sup_{qa}}$ denotes the supervised contrastive loss of myocardium nulling, sharpness, enhancement of aorta and valve, and quality for fibrosis assessment.
After that, we freeze the encoder weights and perform downstream task of supervised learning using the QA module.

\section{Results}

\subsection{Dataset}
Our dataset includes $196$ scans of labeled data for the QA task and $900$ scans that have the blood pool segmentations. All of the scans were acquired as in \cite{BAO:Mar2014}, with a resolution of $1.25\times1.25\times2.5 \text{mm}^{3}$, approximately 15 minutes after gadolinium administration, with a 3D ECG-gated, respiratory navigated gradient echo inversion recovery pulse sequence. The $196$ scans were divided and scored by experts. 
These $196$ scans have a class imbalance problem because most scans are in the $2$ to $4$ range. To address this problem, we have transformed the scores of all attributes, including the fibrosis assessment score, into two different labels: diagnostic and non-diagnostic. Scans with a score of $\geq 3$ are designated as diagnostic, denoted $1$, while less than $3$ is non-diagnostic and denoted as $0$.

\subsection{Data Preprocessing \& Augmentation}
The dataset was split into train, test, and validation sets. The test set contained $20$ patient scans. The remaining scans were divided into training and validation sets in a 90:10 training-to-validation ratio. Each scan was a stack of 2D slices of axial view that was selected by the first stage in Fig. \ref{fig:arch} to contain the left atrium. Images were resized to $128\times128$ using linear interpolation. Since we are assessing image quality, we went for geometric transformations such as Random flip, perspective transform, shift, scale, and rotate, which were applied with a probability of $0.5$ each during training. All data were normalized before being passed through the network.
\begin{figure}[h]
\centering
\includegraphics[width=0.7\textwidth]{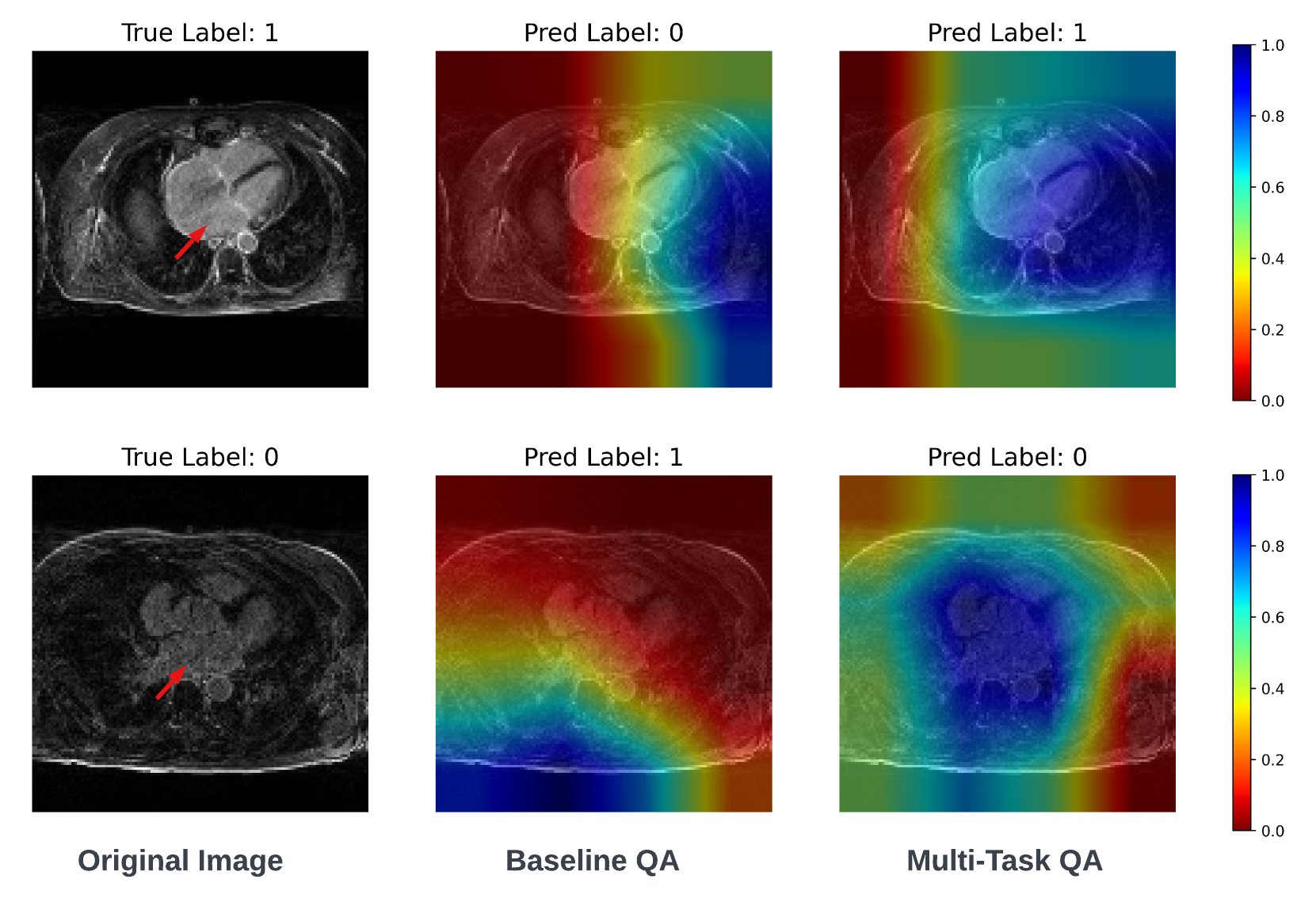}
\caption{Result of HiResCAM \cite{hirescam} between our Baseline QA and Multi-Task QA model output on $2$ different test scans. We show the critical cases where our QA model fails to focus on (blue color) the relevant areas for scoring, whereas, on the contrary, the Multi-Task QA model is able to do so. The "True Label" depicts the ground truth label: diagnostic (1) and non-diagnostic (0), whereas the "Pred Label" denotes the model prediction. A red arrow in the original image shows the location of the left atrium.}
\label{fig:hirescam}
\end{figure}
\begin{figure}[h]
\centering
\includegraphics[width=0.8\textwidth]{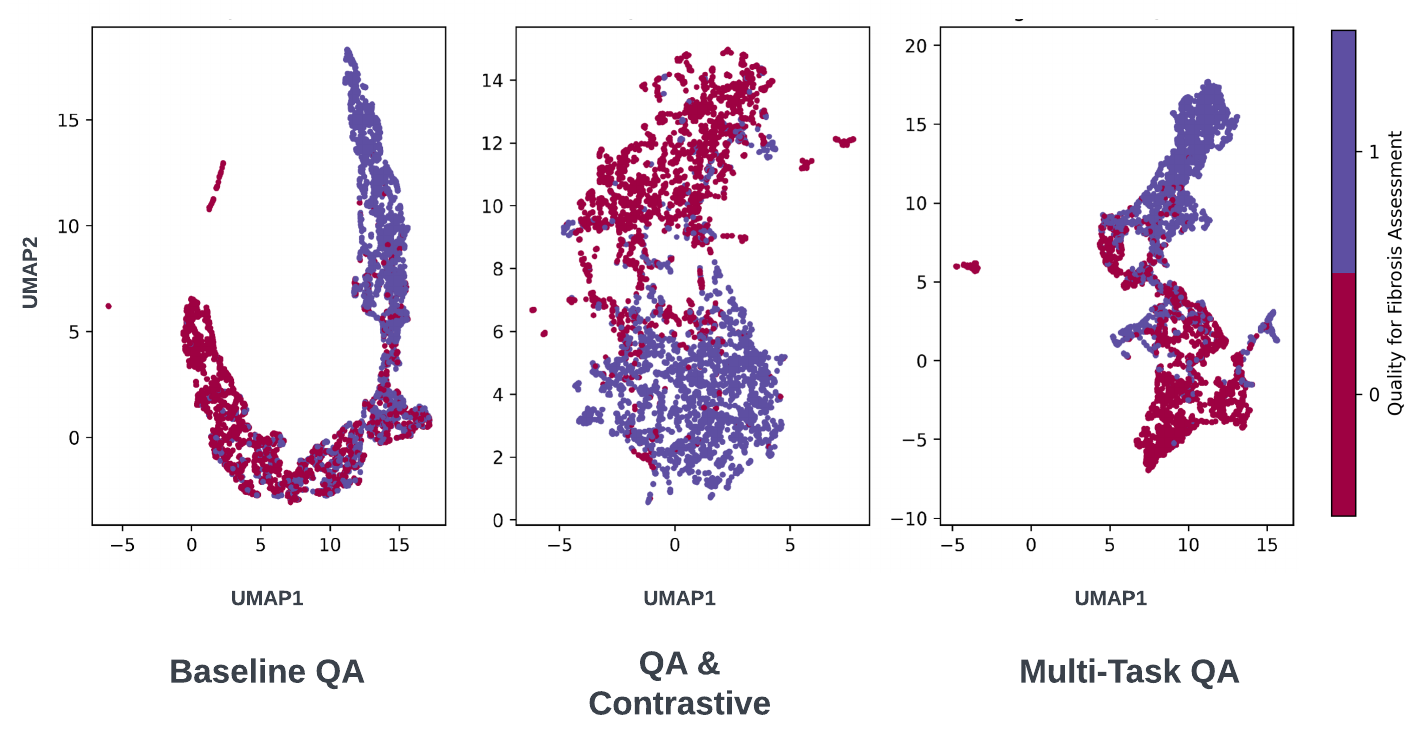}
\caption{UMAP\cite{umap} Visualization of embedding space representation of encoder of the three models. We report the 1st iteration's visualization among the $5$ iterations.}
\label{fig:emb}
\end{figure}

\subsection{Summary of Experiments}
During training, the model is validated using MSE error across all slices on the validation set and later on evaluated on the test set. For the test set, we perform score predictions across all slices and subsequently report the mode for each scan. The performance is measured by Precision, Recall, F1-Score, and Specificity. The details of the training are given below.

\noindent
\textbf{Left Atrium Detection Training:} We train the U-Net network of this stage on the $900$ segmentation masks of the blood pool. The network was trained using the Adam optimizer with a learning rate of $\eta = 0.001$. The batch size was set to $128$. We then use this network to predict the relevant slices of $196$ scans for the later stage.

\noindent
\textbf{Baseline QA Training:} For this model, we only consider the QA module. The network was trained for $40$ epochs using the Adam optimizer with a learning rate of $\eta = 0.001$ and with an early stopping criteria of patience $7$. The batch size was set to $128$. A cosine annealing learning rate scheduler was used to reduce the learning rate throughout training, bringing stable optimization and faster convergence.

\noindent
\textbf{Multi-Task QA Training:} For this model, we consider both the QA and Decoder module in our architecture. The network was trained using the same configuration as the QA Training strategy. To interpret that the Multi-Task QA model focuses on the relevant areas, we report the gradient-based visual explanation from Draelos et al.\cite{hirescam} in Figure \ref{fig:hirescam}.

\noindent
\textbf{QA \& Contrastive Training:} For this model, we follow the 2-step process discussed by Khosla et al.\cite{supconloss}. In the first step, we pre-train the encoder for $100$ epochs with a batch size of $512$, since contrastive learning benefits from more negative samples during training. We use LARS optimizer \cite{lars} with a learning rate of $\eta = 0.5$. In the 2nd step, we freeze the encoder and train the QA module with the same configuration as the QA Training strategy discussed above. Since gradient-based explanation is not suitable for contrastive learning due to the fact that contrastive learning does not require explicit class labels during training, therefore we report the embedding representation by utilizing UMAP\cite{umap} approach presented in Figure \ref{fig:emb}. We can see that the QA \& Contrastive based model formed more tight clusters compared to other methods, thus forming cluster alignment. 
The results of the experiments are shown in Table \ref{table:exp}.

\begin{table}[]
\centering
\resizebox{0.8\textwidth}{!}{
\begin{tabular}{|c|c|c|c|c|c|}
\hline
\textbf{Method}       & \textbf{  Precision  }     & \textbf{  Recall  } & \textbf{  F1-Score  }      & \textbf{Specificity}   \\ \hline
QA                    & $0.60 \pm 0.05$          & $0.90 \pm 0.15$   & $0.710 \pm 0.05$                & $0.38 \pm 0.20$          \\ \hline
Multi-Task QA & ${0.63 \pm 0.07}$ & $0.90 \pm 0.12$   & ${0.73 \pm 0.04}$        & ${0.44 \pm 0.24}$ \\ \hline
QA \& Contrastive & $\mathbf{0.65 \pm 0.07}$ & $\mathbf{0.94 \pm 0.08}$ & $\mathbf{0.76 \pm 0.05}$  & $\mathbf{0.48 \pm 0.17}$ \\ \hline
\end{tabular}}
\caption{The test performance of the networks. We report the means and standard deviations among runs here.}
\label{table:exp}
\end{table}
\section{Conclusion}
In this study, we developed an automated two-stage deep learning model for quality assessment scoring of LGE MRI images using limited labeled data. Our approach includes a segmentation network to identify relevant left atrium slices, and two strategies: Multi-task learning (reconstruction and quality assessment) and supervised contrastive learning (quality assessment). Multi-task learning benefited from segmentation information, improving quality assessment accuracy through knowledge transfer. Supervised contrastive learning effectively learned informative representations from both contrastive learning and label supervision, yielded the most promising results for quality assessment among all other methods. In conclusion, our research provides valuable insights for quality assessment of the left atrium in LGE MRI images with limited labeled data. Hybrid approaches, combining both techniques, could lead to even more robust and accurate quality assessment models for LGE MRI image analysis and other medical imaging domains, addressing data scarcity challenges in medical imaging advancements.

\textbf{Acknowledgements} 
The National Institutes of Health supported this work under grant numbers R01HL162353.

\bibliographystyle{IEEEtran}

\end{document}